\documentclass[conference]{IEEEtran}
\IEEEoverridecommandlockouts
\usepackage{cite}
\usepackage{amsmath,amssymb,amsfonts}
\usepackage{algorithmic, algorithm}
\usepackage{graphicx}
\usepackage{textcomp}
\usepackage{xcolor}
\usepackage{listings}
\usepackage{pgfplots}
\usepackage{pgfplotstable}
\usepackage{tikz-timing}
\usepgfplotslibrary{groupplots}
\usepgfplotslibrary{fillbetween}
\usepackage{multirow}
\usepackage[hidelinks]{hyperref}
\usepackage{soul}
\usepackage{orcidlink}
\def\BibTeX{{\rm B\kern-.05em{\sc i\kern-.025em b}\kern-.08em
    T\kern-.1667em\lower.7ex\hbox{E}\kern-.125emX}}
\usepackage{enumerate}

\newcommand*\circled[1]{\tikz[baseline=(char.base)]{
            \node[shape=circle,fill,inner sep=1pt] (char) {\textcolor{white}{#1}};}}

\pgfplotsset{compat=1.18}

\begin{document}

\title{Advanced Strategies for Uncertainty‑Guided Live Measurement Sequencing in Fast, Robust SAR ADC Linearity Testing}

\author{\IEEEauthorblockN{Thorben Schey\IEEEauthorrefmark{1}\orcidlink{0009-0003-0775-8762},
Khaled Karoonlatifi\IEEEauthorrefmark{2}\orcidlink{0009-0004-7135-6749},
Michael Weyrich\IEEEauthorrefmark{1}\orcidlink{0000-0003-3176-9288}
and Andrey Morozov\IEEEauthorrefmark{1}\orcidlink{0000-0001-6772-8889}}\\
\IEEEauthorblockA{\IEEEauthorrefmark{1}Institute of Industrial Automation and Software Engineering, University of Stuttgart\\
Stuttgart, Germany\\
Email: \{first.last\}@ias.uni-stuttgart.de\\
\IEEEauthorrefmark{2}Advantest Europe GmbH\\
Böblingen, Germany\\
Email: \{first.last\}@advantest.com}}
\maketitle

\IEEEpubid{\begin{minipage}{2.0\columnwidth}
\vspace{50pt}
\centering\footnotesize
© 2025 IEEE. Personal use of this material is permitted. Permission from IEEE must be obtained for all other uses, including reprinting/republishing for advertising or promotional purposes, creating new collective works, for resale or redistribution to servers or lists, or reuse of any copyrighted component of this work in other works.
\end{minipage}}

\begin{abstract}
This paper builds on our Uncertainty-Guided Live Measurement Sequencing (UGLMS) method.
UGLMS is a closed-loop test strategy that adaptively selects SAR ADC code edges based on model uncertainty and refines a behavioral mismatch model in real time via an Extended Kalman Filter (EKF), eliminating full-range sweeps and offline post-processing.
We introduce an enhanced UGLMS that delivers significantly faster test runtimes while maintaining estimation accuracy.
First, a rank-1 EKF update replaces costly matrix inversions with efficient vector operations, and a measurement-aligned covariance-inflation strategy accelerates convergence under unexpected innovations.
Second, we extend the static mismatch model with a low-order carrier polynomial to capture systematic nonlinearities beyond pure capacitor mismatch.
Third, a trace-based termination adapts test length to convergence, preventing premature stops and redundant iterations.
Simulations show the enhanced UGLMS reconstructs full Integral- and Differential-Non-Linearity (INL/DNL) in just 36 ms for 16-bit and under 70 ms for 18-bit ADCs (120 ms with the polynomial extension).
Combining the faster convergence from covariance inflation with reduced per-iteration runtime from the rank-1 EKF update, the method reaches equal accuracy 8× faster for 16-bit ADCs.
These improvements enable real-time, production-ready SAR ADC linearity testing.
\end{abstract}


\section{Introduction}
\label{sec:Introduction}
Analog-to-Digital Converters (ADCs) serve as critical interface components in mixed-signal systems, converting analog sensor or signal voltages into digital codes used in embedded and data-driven applications.
Among various architectures, the Successive Approximation Register (SAR) ADC is widely adopted due to its favorable trade-offs in resolution, power, and area \cite{rapuano2005adc}.
Ensuring the accuracy and reliability of SAR ADCs in production requires thorough linearity testing, typically characterized by Differential Non-Linearity (DNL) and Integral Non-Linearity (INL) \cite{huang2016analysis, vasan2010linearity}.

Conventional test approaches such as histogram-based methods rely on dense input sweeps and post-processing of large datasets, resulting in significant test time and memory demands - especially for high-resolution ADCs \cite{vasan2011adc}.
While model-based and code-selective methods have improved test efficiency, they often require full data acquisition and remain dependent on offline estimation procedures \cite{yu2012algorithm, chaganti2018low}.

To address these limitations, we previously introduced Uncertainty-Guided Live Measurement Sequencing (UGLMS), a closed-loop methodology that adaptively selects measurement points based on model uncertainty and refines a behavioral mismatch model in real time using an Extended Kalman Filter (EKF) \cite{schey2025uncertainty}.
This approach enables accurate INL/DNL estimation without full-code sweeps and without any post-processing, making it highly attractive for integration into production test flows.

The UGLMS method already demonstrated that accurate linearity estimation can be achieved without full code sweeps or post-processing, using only a small number of adaptively chosen measurements.
Building on this foundation, we present targeted refinements that significantly enhance efficiency, flexibility, and robustness in practical test scenarios.

\textbf{Contributions:} We achieve (i) $8\times$ faster equal-accuracy INL/DNL estimation for 16-bit ADCs, (ii) coverage of systematic nonlinearities beyond pure capacitor mismatch, and (iii) automated, convergence-driven test termination.
These gains are realized by (i) a rank-1 EKF update to reduce per-iteration runtime, (ii) a low-order carrier polynomial for extended nonlinearity modeling, and (iii) a measurement-aligned covariance-inflation strategy combined with a trace-based termination criterion for rapid convergence.
We also studied inflation and termination criteria to guide tuning.

The remainder of this paper is organized as follows:
\autoref{sec:State_of_the_Art} surveys ADC linearity testing and model-based estimation strategies.
In \autoref{sec:Methodology}, the original UGLMS method is briefly revisited, followed by a formal introduction of the proposed enhancements.
\autoref{sec:Experimental_Results} presents simulation results and the systematic parameter study.
Finally, \autoref{sec:Conclusion} summarizes the main contributions and outlines future work.

\section{State of the Art}
\label{sec:State_of_the_Art}
Statistical histogram methods have long dominated ADC linearity testing due to their simplicity and robustness.
They characterize Differential Non-Linearity (DNL) - the deviation of each quantization interval from the size of one Least Significant Bit (LSB), where $c$ denotes the ADC output code:
\[
\mathrm{DNL}[c] = \mathrm{CE}[c+1] - \mathrm{CE}[c] - 1.
\]
Integral Non-Linearity (INL) measures the offset of each code edge from its ideal position
\[
\mathrm{INL}[c] = \mathrm{CE}[c] - \mathrm{CE_{ideal}}[c].
\]

In the classic ramp‐based approach, a linear voltage sweep is applied to the ADC input and a code histogram is constructed to estimate DNL and INL \cite{rapuano2005adc,goyal2005test}.
To relax the requirement for an ideal ramp, sine‐wave histogram tests use a spectrally pure tone and reconstruct code densities via the known probability density function of a sine input \cite{rapuano2005adc,huang2016analysis}.
More recent code‐selective variants further reduce measurement effort by concentrating samples on critical regions - such as MSB transition points - yet they still depend on offline histogram processing to derive final linearity metrics \cite{zhao2022evaluation}.

Servo loop techniques place the ADC under test in a feedback configuration with a high accuracy DAC, adjusting the input until the comparator output toggles to pinpoint each code edge \cite{wegener2000model}.
Although this approach yields very precise edge locations, it requires waiting for the feedback loop to settle between measurements.
This settling can take tens of milliseconds and therefore limits throughput when characterizing the full code range.

Offline model-based reconstruction methods build parametric or nonparametric models of the INL curve using far fewer data points.
Segmented approaches such as uSMILE fit locally correlated error segments across MSB, ISB, and LSB levels, reducing samples from millions to thousands \cite{chen2015ultrafast,chaganti2018low}.
Partial-polynomial fitting further approximates the INL curve with low-order polynomials over uneven segments \cite{li2024low,fu2024stimulus}.
Despite this dramatic sample reduction, these methods still depend on full-range data acquisition and offline optimization.

Jin et al. demonstrated that ADC INL estimation can be cast as a Kalman filtering problem, where successive code observations iteratively refine a parameterized model of the code‐edge transfer curve \cite{jin2006linearity}.
Their EKF implementation achieved accurate INL reconstruction with significantly fewer hits per code compared to traditional histograms.
Vasan and Chen extended this concept in their SEIR-KF and KHK-KF methods by embedding stimulus-error identification into the filter update, reducing sample requirements to as low as one hit per code while preserving estimation accuracy \cite{vasan2010linearity}.
Noise filtering inherent to the Kalman method further enhances robustness under measurement variability.
Nevertheless, these methods still depend on offline processing of a complete measurement set and thus do not eliminate the need for full-range data acquisition.

Overall, existing methods either incur high measurement overhead (histogram, servo-loop), rely on full-sweep data acquisition and offline postprocessing (model‐based, Kalman post-processing), or lack adaptability to unexpected error sources.

Uncertainty‐Guided Live Measurement Sequencing (UGLMS) closes the loop between measurement and estimation by selecting code edges based on model uncertainty and updating parameters directly during testing without full sweeps or post‐processing \cite{schey2025uncertainty}.
In Section~\ref{sec:Methodology}, we present UGLMS along with our refinements to this real‐time approach.

\section{Methodology}
\label{sec:Methodology}
\subsection{Baseline UGLMS}
\label{secsub:Baseline_UGLMS}

\begin{figure}[t]
    \setlength{\belowcaptionskip}{-5pt} 
    \centering
    \includegraphics[width=\linewidth]{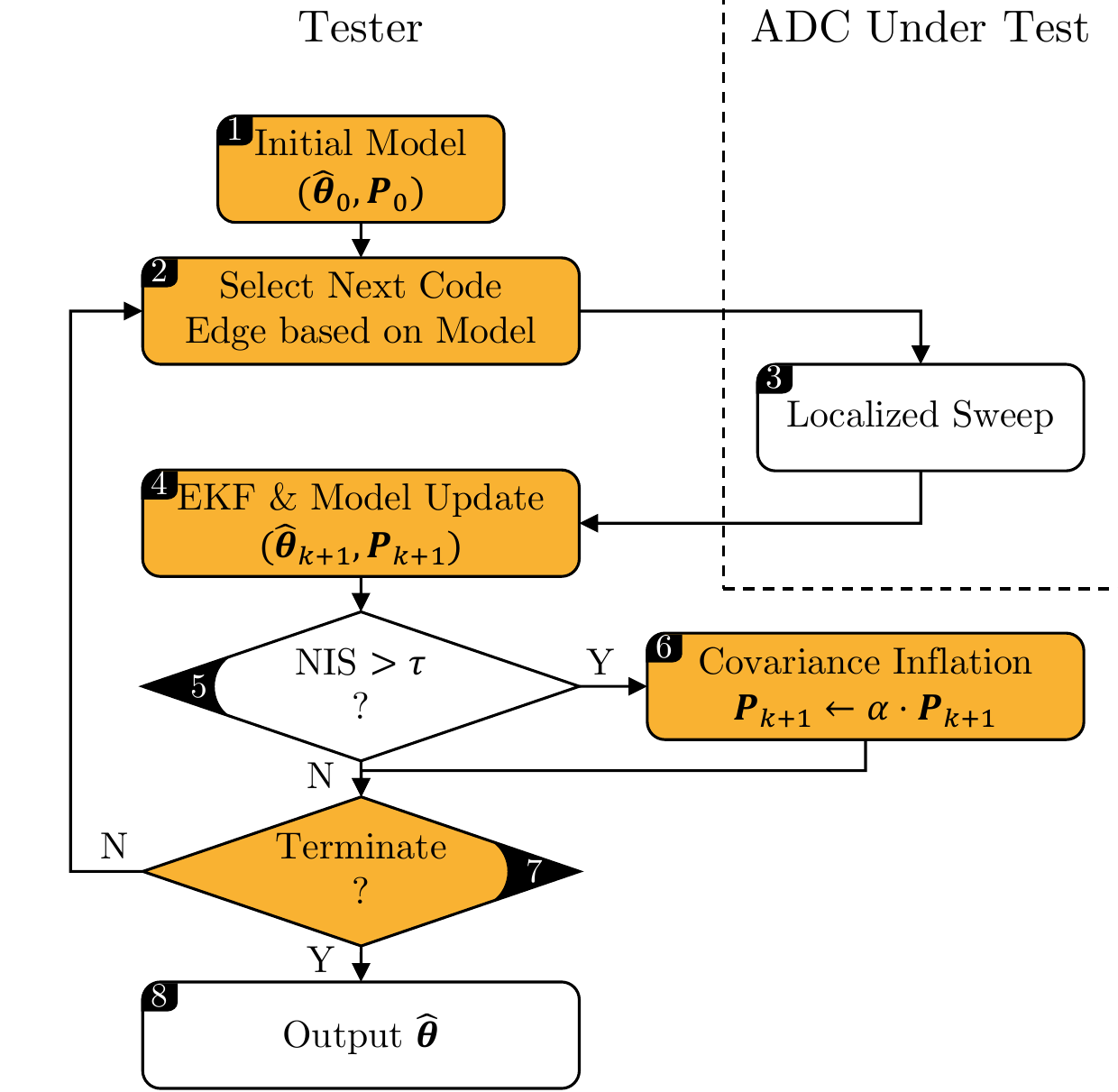}
    \caption{Process flow of the original Uncertainty-Guided Live Measurement Sequencing (UGLMS) for SAR ADC linearity testing \cite{schey2025uncertainty}.
    Highlighted blocks indicate aspects that are adapted and improved in this work.}
    \label{fig:UGLMS_Schematic}
\end{figure}

The baseline Uncertainty-Guided Live Measurement Sequencing (UGLMS) method \cite{schey2025uncertainty} builds on an Extended Kalman Filter (EKF) formulation to adaptively select code edges and update a mismatch parameter estimate $\boldsymbol{\theta}_k$.
It models the ADC transition at code $c$ via a nonlinear measurement function

\[
z_k = f_c(\boldsymbol{\theta}) - f_c(\hat{\boldsymbol{\theta}}_k) + \nu_k,
\quad \nu_k \sim \mathcal{N}(0,R),
\]

Here, $f_c(\boldsymbol{\theta})$ predicts the true analog transition level (in LSB), and $R$ denotes the measurement noise variance. 
The $N \times 1$ gradient (Jacobian) of $f_c$ w.r.t. $\boldsymbol{\theta}$ is denoted by

\[
\mathbf{j}_c = \frac{\partial f_c(\boldsymbol{\theta})}{\partial \boldsymbol{\theta}}\bigg|_{\hat{\boldsymbol{\theta}}_k}.
\]

For a fixed ADC resolution, all per-code gradients $\mathbf{j}_c$ can be computed offline and combined into the Jacobian matrix $\mathbf{H}$.

\autoref{fig:UGLMS_Schematic} visualizes the UGLMS method first proposed in \cite{schey2025uncertainty}.
The UGLMS process continuously interleaves code-edge selection, localized measurement, and recursive updates in an asynchronous update loop:

    \begin{itemize}
    \item[\circled{1}] \textbf{Initialization:} The initial parameter estimate $\boldsymbol{\hat{\theta}}_0$ (e.g. ideal capacitor weights) and covariance $\boldsymbol{P}_0$ (large uncertainties on the diagonal) are set.
    
    \item[\circled{2}] \textbf{Adaptive Code Edge Selection:} For each candidate code $c$, the EKF scalar gain
    \[
    \mathrm{gain}(c) = \frac{\mathbf{j}_c\,\mathbf{P}_k\,\mathbf{j}_c^\top}{\mathbf{j}_c\,\mathbf{P}_k\,\mathbf{j}_c^\top + R}
    \]
    is evaluated, which quantifies the expected reduction in parameter uncertainty by measuring at $c$.
    Then the code edge with the greatest gain is selected:
    \[
    c^* = \arg\max_c\;\mathrm{gain}(c).
    \]
    
    \item[\circled{3}] \textbf{Localized Sweep \& Measurement:} Around the predicted transition level $f_{c^{*}}(\boldsymbol{\hat{\theta}}_k)$, $M$ fine‐step voltages are applied using the tester's high precision DAC (with resolution exceeding that of the ADC under test) and the ADC outputs $y_i$ are recorded.
    The average measured code edge position $\bar{f}_c$ is computed as part of the innovation $z_k$:
    \[
    \bar{f}_c = \frac{1}{M}\sum_{i=1}^M y_i,
    \]
    \[
    z_k = \bar{f}_c - f_{c^*}(\hat{\boldsymbol{\theta}}_k).
    \]

    \item[\circled{4}] \textbf{EKF \& Model Update:} Using the standard extended Kalman update the parameter estimate and covariance are refined:
    \[
    \mathbf{K}_k = \frac{\mathbf{P}_k\,\mathbf{j}_{c^*}^\top}{\mathbf{j}_{c^*}\,\mathbf{P}_k\,\mathbf{j}_{c^*}^\top + R},
    \]
    \[
    \hat{\boldsymbol{\theta}}_{k+1} = \hat{\boldsymbol{\theta}}_k + \mathbf{K}_k\,z_k,
    \]
    \[
    \mathbf{P}_{k+1} = (\mathbf{I} - \mathbf{K}_k\,\mathbf{j}_{c^*})\,\mathbf{P}_k.
    \]
    
    \item[\circled{5}] \textbf{Normalized Innovation Squared (NIS):} The NIS metric is computed to detect situations where the observed measurement deviation exceeds its predicted uncertainty (with $S_k$ being the measurement variance):
    \[
    S_k = \mathbf{j}_{c^*}\,\mathbf{P}_k\,\mathbf{j}_{c^*}^\top + R,
    \]
    \[
    \mathrm{NIS}_k = \frac{z_k^2}{S_k}.
    \]
    
    \item[\circled{6}] \textbf{Covariance Inflation:} If $\mathrm{NIS}_k > \tau$, the updated covariance is inflated uniformly:
    \[
    \mathbf{P}_{k+1} \gets \alpha \cdot \mathbf{P}_{k+1}.
    \]

    \item[\circled{7}] \textbf{Termination Criterion:} Exit the loop once a predefined convergence condition is met, such as a maximum iteration count, small covariance trace, low innovation magnitude, or elapsed time limit.

    
    \item[\circled{8}] \textbf{Termination:} Once the termination criterion is satisfied, output the final estimate $\boldsymbol{\hat\theta}$.
    The INL/DNL curves can then be reconstructed efficiently from $\boldsymbol{\hat\theta}$.
    
    \end{itemize}

\subsection{Rank-1 EKF update and gain selection}
\label{secsub:Rank1_EKF}
In many EKF implementations, both the state covariance update and the per‑code gain computation involve dense operations on the \(N\times N\) covariance matrix, leading to \(\mathcal O(N^2)\) work per measurement. 
By exploiting the fact that each measurement is scalar, these updates can be cast in a rank‑1 form that avoids any matrix inversion and replaces dense matrix-matrix products with matrix-vector and outer products.

First, the covariance update becomes
\[
\mathbf{P}_{k+1} = \mathbf{P}_k - \frac{\bigl(\mathbf{P}_k\,\mathbf{j}_{c^*}^\top \bigr) \, \bigl(\mathbf{P}_k\,\mathbf{j}_{c^*}^\top \bigr)^\top}{\mathbf{j}_{c^*}\,\mathbf{P}_k\,\mathbf{j}_{c^*}^\top + R},
\]
which requires exactly two matrix-vector products and one outer product (total \(\mathcal O(N^2)\)) but no matrix inversion.

For code‑edge selection, recomputing
\(\mathbf{j}_c\,\mathbf{P}_k\,\mathbf{j}_c^\top\)
for all \(C=2^{n_{\rm bits}}\) edges would cost \(\mathcal O(C\,N^2)\).
Instead, a GPU‑resident array
\[
\mathbf{Q}_k[c] = \mathbf{j}_c\,\mathbf{P}_k\,\mathbf{j}_c^\top
\]
is kept in memory and the same rank‑1 correction is applied after measuring at \(s=c^*\):
\[
\mathbf{Q}_{k+1}[c] = \mathbf{Q}_k[c] - \frac{\bigl(\mathbf{j}_c\,\mathbf{P}_k\,\mathbf{j}_s^\top\bigr)^2} {\mathbf{j}_s\,\mathbf{P}_k\,\mathbf{j}_s^\top + R}.
\]

On the GPU this is implemented via four parallel kernels:
\begin{enumerate}
  \item Compute \(\mathbf{v} = \mathbf{P}_k\,\mathbf{j}_s^\top\) \hfill \(\mathcal O(N^2)\)
  \item Compute \(\mathbf{u}[c] = \mathbf{j}_c\,\mathbf{v}\) for all \(c\) \hfill \(\mathcal O(C\,N)\)
  \item Elementwise update  
    \(\mathbf{Q}[c] \gets \mathbf{Q}[c] - \tfrac{\mathbf{u}[c]^2}{\mathbf{Q}_k[s] + R}\) \hfill \(\mathcal O(C)\)
  \item Apply the rank‑1 covariance update to \(\mathbf{P}_k\) \hfill \(\mathcal O(N^2)\)
\end{enumerate}
Overall, this reduces the per-measurement complexity from $\mathcal O(CN^2)$ to $\mathcal O(N^2+CN)$, confining the dominant computation to just two $N^2$ matrix-vector products and a linear-cost update over all code edges.

\subsection{Non-uniform covariance inflation}
\label{secsub:CovarianceInflation}
The uniform scaling of \(\mathbf{P}_{k+1}\) in the baseline UGLMS method can over‑inflate directions orthogonal to the recent measurement and lead to suboptimal convergence behavior.
To confine the added uncertainty to the measured subspace, measurement‑aligned inflation is applied along the latest measurement direction:
\[
\mathbf{P}_{k+1} \gets \mathbf{P}_{k+1} + (\alpha-1)\,
    \frac{\bigl(\mathbf{P}_k\,\mathbf{j}_{c^*}^\top\bigr)\,
    \bigl(\mathbf{P}_k\,\mathbf{j}_{c^*}^\top\bigr)^\top}
    {\mathbf{j}_{c^*}\,\mathbf{P}_k\,\mathbf{j}_{c^*}^\top}\,.
\]

Here \(\mathbf{P}_{k+1}\) on the left is the post‑update covariance before inflation.
To reduce the \(\mathcal O(N^2)\) cost of a full rank‑1 update, we approximate by updating only the diagonal:
\[
\mathbf{P}_{k+1}[i,i] \gets \mathbf{P}_{k+1}[i,i] + (\alpha-1)\,\frac{\bigl(\mathbf{P}_k[i]\,\mathbf{j}_{c^*}^\top\bigr)^{2}} {\mathbf{j}_{c^*}\,\mathbf{P}_k\,\mathbf{j}_{c^*}^\top}.
\]
This diagonal update runs in \(\mathcal O(N)\) and concentrates added variance along the direction of the latest measurement, avoiding equal inflation of partially or fully unrelated subspaces.

\begin{table*}[!t]
    \centering
    \caption{
    Test time comparison for SAR ADCs with different resolutions for different UGLMS variants.
    The code selection time refers to the computational effort of selecting the next code to test.
    All tests executed 200 iterations with input noise of $1.0\,\mathrm{LSB}$ RMS.
    }
    \label{tab:test_times}
    
    \begin{tabular}{cc | rr |  rr | rr}
        \hline
        & & \multicolumn{2}{c|}{Baseline \cite{schey2025uncertainty}} 
          & \multicolumn{2}{c|}{Rank-1} 
          & \multicolumn{2}{c}{Rank-1 + Poly} \\
        Resolution $N$ & Samples/Sweep $M$
          & Select [$\mu \mathrm{s}$] & Total [$\mathrm{ms}$]
          & Select [$\mu \mathrm{s}$] & Total [$\mathrm{ms}$] 
          & Select [$\mu \mathrm{s}$] & Total [$\mathrm{ms}$] \\
        \hline
        $10$ & \makebox[1.5em][r]{$64$} & $88.0$ & $18.5$ & $106.0$ \textsubscript{(\makebox[2.5em][r]{\phantom{0}+20\%})} & $21.6$ \textsubscript{(\makebox[2.5em][r]{\phantom{0}+17\%})} & $114.0$ \textsubscript{(\makebox[2.5em][r]{\phantom{0}+30\%})} & $23.4$ \textsubscript{(\makebox[2.5em][r]{\phantom{0}+26\%})} \\
        $12$ & \makebox[1.5em][r]{$64$} & $89.5$ & $18.7$ & $113.0$ \textsubscript{(\makebox[2.5em][r]{\phantom{0}+26\%})} & $23.8$ \textsubscript{(\makebox[2.5em][r]{\phantom{0}+27\%})} & $115.0$ \textsubscript{(\makebox[2.5em][r]{\phantom{0}+28\%})} & $24.5$ \textsubscript{(\makebox[2.5em][r]{\phantom{0}+31\%})} \\
        $14$ & $128$ & $141.0$ & $31.6$ & $116.0$ \textsubscript{(\makebox[2.5em][r]{\phantom{0}-18\%})} & $24.4$ \textsubscript{(\makebox[2.5em][r]{\phantom{0}-23\%})} & $117.0$ \textsubscript{(\makebox[2.5em][r]{\phantom{0}-17\%})} & $27.3$ \textsubscript{(\makebox[2.5em][r]{\phantom{0}-14\%})} \\
        $16$ & $128$ & $300.5$ & $61.2$ & $149.0$ \textsubscript{(\makebox[2.5em][r]{\phantom{0}-50\%})} & $36.0$ \textsubscript{(\makebox[2.5em][r]{\phantom{0}-41\%})} & $160.0$ \textsubscript{(\makebox[2.5em][r]{\phantom{0}-47\%})} & $42.3$ \textsubscript{(\makebox[2.5em][r]{\phantom{0}-31\%})} \\
        $18$ & $128$ & $1181.0$ & $238.8$ & $308.0$ \textsubscript{(\makebox[2.5em][r]{\phantom{0}-74\%})} & $67.1$ \textsubscript{(\makebox[2.5em][r]{\phantom{0}-72\%})} & $418.0$ \textsubscript{(\makebox[2.5em][r]{\phantom{0}-65\%})} & $120.1$ \textsubscript{(\makebox[2.5em][r]{\phantom{0}-50\%})} \\
        \hline
    \end{tabular}
\end{table*}

\subsection{Robust termination strategy}
\label{secsub:TerminationStrategy}
A simple threshold on $\mathrm{trace}(\mathbf{P}_k)$ can lead to premature termination, as covariance inflation may temporarily increase the trace despite ongoing estimation progress.
To improve robustness, a smoothed criterion based on recent trace evolution is proposed: the loop terminates only if the difference
\[
\Delta_k = \mathrm{trace}(\mathbf{P}_k) - \mathrm{trace}(\mathbf{P}_{k-1})
\]
remains below a threshold $\varepsilon$ for $N_\mathrm{term}$ consecutive iterations, thereby suppressing false convergence triggers.
The value of $\varepsilon$ is subject to further tuning, as analyzed in \autoref{secsub:Results_Epsilon}.

\subsection{Carrier polynomial extension}
\label{secsub:CarrierPoly}
To capture residual systematic nonlinearities (e.g. comparator drift), the ADC model is extended to include a low‑order polynomial of order $o$ in the code index $c$:
\[
h(\boldsymbol\theta;c)
= f_c(\boldsymbol\theta_{\rm mismatch})
  \;+\;
  \sum_{m=0}^{\,o}\beta_m\,c^m.
\]
Accordingly, the state vector is expanded to
\(\displaystyle \widetilde{\boldsymbol\theta}
  = \bigl[\boldsymbol\theta_{\rm mismatch}^\top,\;\beta_{0},\dots,\beta_{o}\bigr]^\top,\)
and each Jacobian row becomes
\[
\mathbf{j}_c
= \bigl[\,
  \tfrac{\partial f_c}{\partial\boldsymbol\theta_{\rm mismatch}}\,,\;
  c^{0},\,c^{1},\,\dots,\,c^{o}
  \bigr].
\]
Both the adaptive code‑edge selection and the rank‑1 EKF update (including covariance and gain adjustments) extend directly to this $(N+o+1)$-dimensional state, preserving the same live‑loop structure.

\section{Experimental results and parameter study}
\label{sec:Experimental_Results}
To assess the impact of the proposed improvements and investigate key tuning parameters, simulation experiments were conducted.
The behavior of the SAR ADC is modeled using mismatch vectors obtained from measured devices.
To preserve confidentiality, these vectors were resampled and obfuscated to maintain their statistical properties.
All algorithms and test environments were implemented in C for computational efficiency.

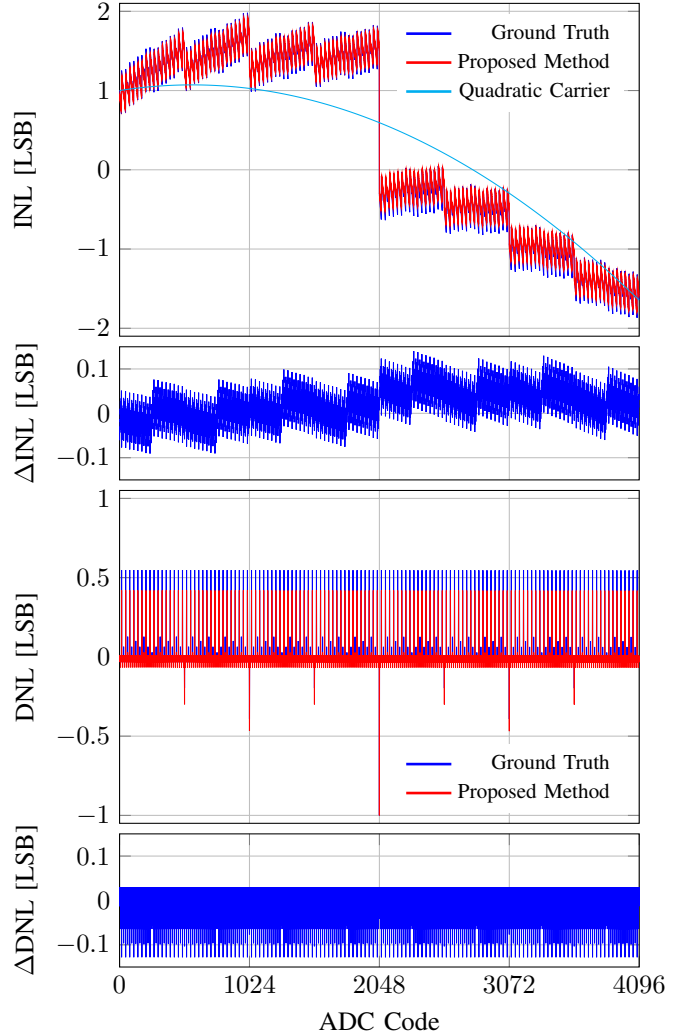
\begin{figure}[t]
    \setlength{\abovecaptionskip}{-5pt} 
    \setlength{\belowcaptionskip}{-5pt} 
    \centering
    \begin{tikzpicture}
\begin{groupplot}[
    group style={
        group size=1 by 4,
        vertical sep=4pt,
        horizontal sep=0cm,
    },
    width=0.78\linewidth,
    height=0.5\linewidth,
    scale only axis=true,
    xmin=0, xmax=4096,
    grid=major,
    legend style={draw=none, font=\footnotesize},
    legend cell align={right},
    legend image post style={line width=1pt},
]

\nextgroupplot[
    ylabel={INL [LSB]},
    ylabel style={yshift=8.5pt},
    ymin=-2.1, ymax=2.1,
    xtick distance=1024,
    xmajorticks=false,
    legend pos=north east,
]
\addplot[blue] table {Plots/SingleRun_Group2/true_inl.dat};
\addlegendentry{Ground Truth};
\addplot[red] table {Plots/SingleRun_Group2/est_inl.dat};
\addlegendentry{Proposed Method};
\addplot[cyan] table {Plots/SingleRun_Group2/poly_inl.dat};
\addlegendentry{Quadratic Carrier};

\nextgroupplot[
    ylabel={$ \Delta $INL [LSB]},
    ymin=-0.15, ymax=0.15,
    xmajorticks=false,
    xtick distance=1024,
    ytick distance=0.1,
    height=0.2\linewidth,
]
\addplot[blue] table {Plots/SingleRun_Group2/diff_inl.dat};

\nextgroupplot[
    ylabel={DNL [LSB]},
    ymin=-1.05, ymax=1.05,
    xtick distance=1024,
    xmajorticks=false,
    legend pos=south east,
]
\addplot[blue] table {Plots/SingleRun_Group2/true_dnl.dat};
\addlegendentry{Ground Truth};
\addplot[red] table {Plots/SingleRun_Group2/est_dnl.dat};
\addlegendentry{Proposed Method};

\nextgroupplot[
    xlabel={ADC Code},
    ylabel={$ \Delta $DNL [LSB]},
    ymin=-0.15, ymax=0.15,
    xtick distance=1024,
    xticklabel style={/pgf/number format/1000 sep=},
    ytick distance=0.1,
    height=0.2\linewidth,
]
\addplot[blue] table {Plots/SingleRun_Group2/diff_dnl.dat};

\end{groupplot}

\end{tikzpicture}
    \caption{
    INL and DNL of a 12-bit ADC reconstructed after $200$ iterations with $M=64$ samples per local sweep and $1.0\,\mathrm{LSB}$ RMS measurement noise.
    A quadratic carrier term introduces a $1\,\mathrm{LSB}$ offset and non-uniform gain to model systematic nonlinearity.
    Each estimated curve (INL, DNL) is followed by its corresponding absolute error.
    }
    \label{fig:Group_INL_DNL}
\end{figure}

\subsection{Test time improvements via rank-1 update}
\label{secsub:Results_Time}
\autoref{tab:test_times} shows the per code selection latency and total test time for three UGLMS variants: the original baseline, the rank-1 EKF update with directional inflation, and that same configuration extended by a quadratic carrier term, measured across ADC resolutions from 10 to 18 bits.

For a 16-bit ADC, the per code selection step drops from $300.5\,\mu \mathrm{s}$ to $149.0\,\mu \mathrm{s}$ ($2.0\times$) and total test time falls from $61.2\,\mathrm{ms}$ to $36.0\,\mathrm{ms}$.
At 18 bits the selection time improves from $1181\,\mu \mathrm{s}$ to $308\,\mu \mathrm{s}$ ($3.8\times$), reducing runtime from $238.8\,\mathrm{ms}$ to $67.1\,\mathrm{ms}$.

\newpage
For lower resolutions such as 10 and 12 bits, per code selection latency increases due to fixed overheads, which dominate the modest savings in vector operations.
Consequently, the original UGLMS implementation remains more efficient at these lower resolutions.

The adapted inflation strategy does not increase selection latency as it only operates on the covariance diagonal.
In contrast, adding the carrier polynomial introduces extra state dimensions and Jacobian entries, which increases both selection and total runtime.
This can be seen as a reasonable trade-off for the enhanced modeling capability.

\subsection{Model extension with carrier polynomial}
\label{secsub:Results_Poly}
A quadratic carrier term was added to the capacitor‐mismatch model and evaluated on a 12-bit ADC with synthetic second-order distortion.
After 200 iterations, the estimated INL and DNL curves closely align with the true INL and DNL curves (see \autoref{fig:Group_INL_DNL}), including the overlaid quadratic distortion, with deviations limited to $\pm 0.15\,\mathrm{LSB}$.
The model successfully inherits both the $+1\,\mathrm{LSB}$ offset and the nonlinear gain introduced by the synthetic polynomial, thereby confirming the extension’s ability to capture systematic nonlinearity beyond capacitor mismatch.

\begin{figure}[t]
    \setlength{\abovecaptionskip}{-5pt} 
    \begin{tikzpicture}
\begin{axis}[
    ylabel={$\Delta$INL\textsubscript{max} [LSB]},
    grid=major,
    ymin=0,
    ymax=0.5,
    xmin=0,
    xmax=1000,
    xtick distance=100,
    xticklabel style={/pgf/number format/1000 sep=},
    xmajorticks=false,
    width=\linewidth,
    height=0.525\linewidth,
    legend pos=north east,
    legend cell align={left},
    legend style={draw=none, font=\footnotesize, xshift=6pt, yshift=2.5pt},
    legend image post style={line width=1pt},
]
\addplot[red!30, forget plot, name path=A] table {Plots/VarIter_Innovation/p90_inl_new.dat};
\addplot[red!30, forget plot, name path=B] table {Plots/VarIter_Innovation/p10_inl_new.dat};
\addplot[red!10, forget plot, fill opacity=0.6] fill between[of=A and B];

\addplot[blue!30, forget plot, name path=C] table {Plots/VarIter_Innovation/p90_inl_old.dat};
\addplot[blue!30, forget plot, name path=D] table {Plots/VarIter_Innovation/p10_inl_old.dat};
\addplot[blue!10, forget plot, fill opacity=0.6] fill between[of=C and D];

\addplot[red, thick] table {Plots/VarIter_Innovation/mean_inl_new.dat};
\addlegendentry{Measurement-aligned}

\addplot[blue, thick] table {Plots/VarIter_Innovation/mean_inl_old.dat};
\addlegendentry{Uniform}
\end{axis}
\end{tikzpicture}

\begin{tikzpicture}
\begin{axis}[
    xlabel={Iteration},
    ylabel={$\Delta$DNL\textsubscript{max} [LSB]},
    grid=major,
    ymin=0,
    ymax=0.5,
    xmin=0,
    xmax=1000,
    xtick distance=100,
    xticklabel style={/pgf/number format/1000 sep=},
    width=\linewidth,
    height=0.525\linewidth,
    legend pos=north east,
    legend cell align={left},
    legend style={draw=none, font=\footnotesize, xshift=6pt, yshift=2.5pt},
    legend image post style={line width=1pt},
]
\addplot[red!30, forget plot, name path=A] table {Plots/VarIter_Innovation/p90_dnl_new.dat};
\addplot[red!30, forget plot, name path=B] table {Plots/VarIter_Innovation/p10_dnl_new.dat};
\addplot[red!10, forget plot, fill opacity=0.6] fill between[of=A and B];

\addplot[blue!30, forget plot, name path=C] table {Plots/VarIter_Innovation/p90_dnl_old.dat};
\addplot[blue!30, forget plot, name path=D] table {Plots/VarIter_Innovation/p10_dnl_old.dat};
\addplot[blue!10, forget plot, fill opacity=0.6] fill between[of=C and D];

\addplot[red, thick] table {Plots/VarIter_Innovation/mean_dnl_new.dat};
\addlegendentry{Measurement-aligned}

\addplot[blue, thick] table {Plots/VarIter_Innovation/mean_dnl_old.dat};
\addlegendentry{Uniform}
\end{axis}
\end{tikzpicture}
    \caption{
    Convergence of INL\textsubscript{max} and DNL\textsubscript{max} estimation error over $1000$ iterations for a 16-bit ADC with $1.0\,\mathrm{LSB}$ RMS measurement noise and $M=128$ samples per sweep. Both the uniform inflation method from \cite{schey2025uncertainty} and the proposed measurement-aligned inflation were evaluated across $100$ runs. Mean error is shown with shaded envelopes marking the 10th and 90th percentiles.
    }
    \label{fig:VarIter_Innovation}
\end{figure}
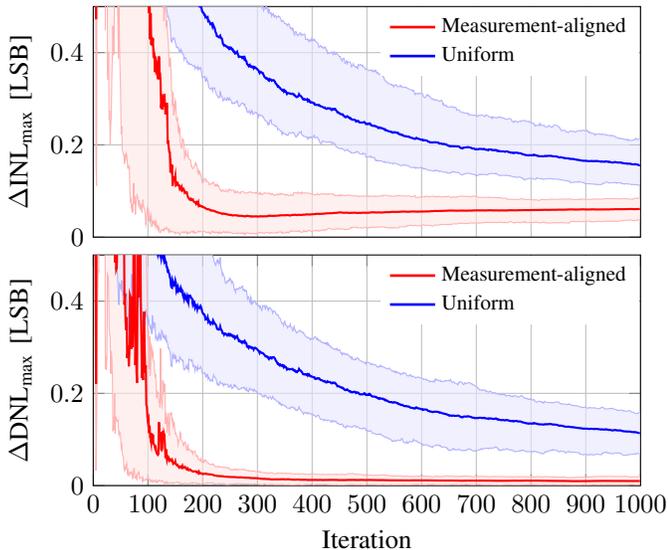

\subsection{Convergence analysis and covariance inflation study}
\label{secsub:Results_Inflation}
\autoref{fig:VarIter_Innovation} compares the convergence behavior of the original uniform inflation strategy and the proposed measurement-aligned variant introduced in~\autoref{secsub:CovarianceInflation}.
The new approach leads to faster convergence and shows less variance across different runs.

\begin{figure}[t]
    \setlength{\abovecaptionskip}{-10pt} 
    \setlength{\belowcaptionskip}{-10pt} 
    \input{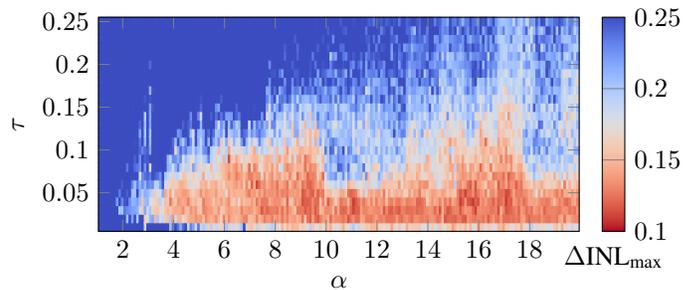}
    \caption{
    INL\textsubscript{max} estimation error as a function of the NIS threshold $\tau$ and inflation factor $\alpha$ for a 16-bit ADC with $M=128$ samples per sweep and $1.0\,\mathrm{LSB}$ RMS noise.
    A total of 4725 parameter combinations ($\tau \in [0.01,0.25]$, $\alpha \in [1.1,20.0]$) were evaluated, each averaged over 20 independent simulation runs.
    The colormap spans a focused error range of $0.1-0.25\,\mathrm{LSB}$ to highlight regions of interest; deep blue indicates configurations with errors $\ge 0.25\,\mathrm{LSB}$.
    }
    \label{fig:Heatmap}
\end{figure}

\autoref{fig:Heatmap} visualizes the impact of different inflation settings.
A general tendency toward small $\tau$ values in the range $0.02-0.06$ is observed for all $\alpha > 3.8$.
Two broad $\alpha$ regions yield particularly low INL errors: $6-9.5$ and $15-17.5$.
Among these, $(\alpha, \tau) = (9.1, 0.04)$ is chosen as a recommended setting.
It yields one of the lowest INL deviations while lying in a broad, stable region within the lower $\alpha$ range, making it robust to parameter variation without excessive inflation.

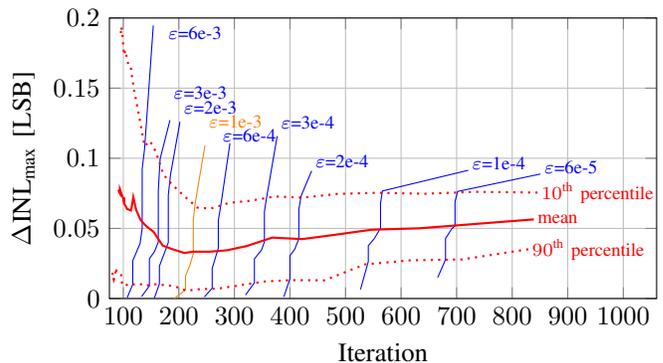
\begin{figure}[t]
    \begin{tikzpicture}
\begin{axis}[
    xlabel={Iteration},
    ylabel={$\Delta$INL\textsubscript{max} [LSB]},
    grid=major,
    ymin=0,
    ymax=0.2,
    xmin=75,
    xmax=1060,
    xtick distance=100,
    xticklabel style={/pgf/number format/1000 sep=},
    yticklabel style={/pgf/number format/fixed, /pgf/number format/precision=3},
    width=\linewidth,
    height=0.6\linewidth,
]
\addplot[blue, thin] table {Plots/Epsilon/fan_export/fan_eps_1em04.dat};
\addplot[blue, thin] table {Plots/Epsilon/fan_export/fan_eps_2em03.dat};
\addplot[blue, thin] table {Plots/Epsilon/fan_export/fan_eps_2em04.dat};
\addplot[blue, thin] table {Plots/Epsilon/fan_export/fan_eps_3em03.dat};
\addplot[blue, thin] table {Plots/Epsilon/fan_export/fan_eps_3em04.dat};
\addplot[blue, thin] table {Plots/Epsilon/fan_export/fan_eps_6em03.dat};
\addplot[blue, thin] table {Plots/Epsilon/fan_export/fan_eps_6em04.dat};
\addplot[blue, thin] table {Plots/Epsilon/fan_export/fan_eps_6em05.dat};

\addplot[orange, thin] table {Plots/Epsilon/fan_export/fan_eps_1em03.dat};

\addplot[red, thick, dotted] table {Plots/Epsilon/fan_export/line_10p.dat};

\addplot[red, thick, solid] table {Plots/Epsilon/fan_export/line_mean.dat};

\addplot[red, thick, dotted] table {Plots/Epsilon/fan_export/line_90p.dat};

\pgfplotstableread{Plots/Epsilon/fan_export/fan_eps_1em03.dat}\datatableA
\pgfplotstablegetrowsof{\datatableA}
\pgfmathtruncatemacro{\lastrow}{\pgfplotsretval - 1} 
\pgfplotstablegetelem{\lastrow}{[index]0}\of{\datatableA}  
\let\xlastA\pgfplotsretval
\pgfplotstablegetelem{\lastrow}{[index]1}\of{\datatableA}  
\let\ylastA\pgfplotsretval
\node[orange, anchor=west, font=\scriptsize] at (axis cs:\xlastA-10, \ylastA+0.017) {$\varepsilon$=1e-3};

\pgfplotstableread{Plots/Epsilon/fan_export/fan_eps_1em04.dat}\datatableB
\pgfplotstablegetrowsof{\datatableB}
\pgfmathtruncatemacro{\lastrow}{\pgfplotsretval - 1} 
\pgfplotstablegetelem{\lastrow}{[index]0}\of{\datatableB}  
\let\xlastB\pgfplotsretval
\pgfplotstablegetelem{\lastrow}{[index]1}\of{\datatableB}  
\let\ylastB\pgfplotsretval
\node[blue, anchor=west, font=\scriptsize] at (axis cs:\xlastB-10, \ylastB+0.005) {$\varepsilon$=1e-4};

\pgfplotstableread{Plots/Epsilon/fan_export/fan_eps_2em03.dat}\datatableC
\pgfplotstablegetrowsof{\datatableC}
\pgfmathtruncatemacro{\lastrow}{\pgfplotsretval - 1} 
\pgfplotstablegetelem{\lastrow}{[index]0}\of{\datatableC}  
\let\xlastC\pgfplotsretval
\pgfplotstablegetelem{\lastrow}{[index]1}\of{\datatableC}  
\let\ylastC\pgfplotsretval
\node[blue, anchor=west, font=\scriptsize] at (axis cs:\xlastC-10, \ylastC+0.010) {$\varepsilon$=2e-3};

\pgfplotstableread{Plots/Epsilon/fan_export/fan_eps_2em04.dat}\datatableD
\pgfplotstablegetrowsof{\datatableD}
\pgfmathtruncatemacro{\lastrow}{\pgfplotsretval - 1} 
\pgfplotstablegetelem{\lastrow}{[index]0}\of{\datatableD}  
\let\xlastD\pgfplotsretval
\pgfplotstablegetelem{\lastrow}{[index]1}\of{\datatableD}  
\let\ylastD\pgfplotsretval
\node[blue, anchor=west, font=\scriptsize] at (axis cs:\xlastD-10, \ylastD+0.007) {$\varepsilon$=2e-4};

\pgfplotstableread{Plots/Epsilon/fan_export/fan_eps_3em03.dat}\datatableE
\pgfplotstablegetrowsof{\datatableE}
\pgfmathtruncatemacro{\lastrow}{\pgfplotsretval - 1} 
\pgfplotstablegetelem{\lastrow}{[index]0}\of{\datatableE}  
\let\xlastE\pgfplotsretval
\pgfplotstablegetelem{\lastrow}{[index]1}\of{\datatableE}  
\let\ylastE\pgfplotsretval
\node[blue, anchor=west, font=\scriptsize] at (axis cs:\xlastE-10, \ylastE+0.020) {$\varepsilon$=3e-3};

\pgfplotstableread{Plots/Epsilon/fan_export/fan_eps_3em04.dat}\datatableF
\pgfplotstablegetrowsof{\datatableF}
\pgfmathtruncatemacro{\lastrow}{\pgfplotsretval - 1} 
\pgfplotstablegetelem{\lastrow}{[index]0}\of{\datatableF}  
\let\xlastF\pgfplotsretval
\pgfplotstablegetelem{\lastrow}{[index]1}\of{\datatableF}  
\let\ylastF\pgfplotsretval
\node[blue, anchor=west, font=\scriptsize] at (axis cs:\xlastF-10, \ylastF+0.010) {$\varepsilon$=3e-4};

\pgfplotstableread{Plots/Epsilon/fan_export/fan_eps_6em03.dat}\datatableG
\pgfplotstablegetrowsof{\datatableG}
\pgfmathtruncatemacro{\lastrow}{\pgfplotsretval - 1} 
\pgfplotstablegetelem{\lastrow}{[index]0}\of{\datatableG}  
\let\xlastG\pgfplotsretval
\pgfplotstablegetelem{\lastrow}{[index]1}\of{\datatableG}  
\let\ylastG\pgfplotsretval
\node[blue, anchor=west, font=\scriptsize] at (axis cs:\xlastG+10, \ylastG-0.005) {$\varepsilon$=6e-3};

\pgfplotstableread{Plots/Epsilon/fan_export/fan_eps_6em04.dat}\datatableH
\pgfplotstablegetrowsof{\datatableH}
\pgfmathtruncatemacro{\lastrow}{\pgfplotsretval - 1} 
\pgfplotstablegetelem{\lastrow}{[index]0}\of{\datatableH}  
\let\xlastH\pgfplotsretval
\pgfplotstablegetelem{\lastrow}{[index]1}\of{\datatableH}  
\let\ylastH\pgfplotsretval
\node[blue, anchor=west, font=\scriptsize] at (axis cs:\xlastH-30, \ylastH+0.006) {$\varepsilon$=6e-4};

\pgfplotstableread{Plots/Epsilon/fan_export/fan_eps_6em05.dat}\datatableI
\pgfplotstablegetrowsof{\datatableI}
\pgfmathtruncatemacro{\lastrow}{\pgfplotsretval - 1} 
\pgfplotstablegetelem{\lastrow}{[index]0}\of{\datatableI}  
\let\xlastI\pgfplotsretval
\pgfplotstablegetelem{\lastrow}{[index]1}\of{\datatableI}  
\let\ylastI\pgfplotsretval
\node[blue, anchor=west, font=\scriptsize] at (axis cs:\xlastI-10, \ylastI+0.005) {$\varepsilon$=6e-5};


\pgfplotstableread{Plots/Epsilon/fan_export/line_10p.dat}\datatableL
\pgfplotstablegetelem{0}{[index]0}\of{\datatableL}  
\let\xfirstL\pgfplotsretval
\pgfplotstablegetelem{0}{[index]1}\of{\datatableL}  
\let\yfirstL\pgfplotsretval
\node[red, anchor=west, font=\scriptsize] at (axis cs:\xfirstL-10, \yfirstL) {$90^{\text{th}}$ percentile};

\pgfplotstableread{Plots/Epsilon/fan_export/line_90p.dat}\datatableU
\pgfplotstablegetelem{0}{[index]0}\of{\datatableU}  
\let\xfirstU\pgfplotsretval
\pgfplotstablegetelem{0}{[index]1}\of{\datatableU}  
\let\yfirstU\pgfplotsretval
\node[red, anchor=west, font=\scriptsize] at (axis cs:\xfirstU-10, \yfirstU) {$10^{\text{th}}$ percentile};

\pgfplotstableread{Plots/Epsilon/fan_export/line_mean.dat}\datatableM
\pgfplotstablegetelem{0}{[index]0}\of{\datatableM}  
\let\xfirstM\pgfplotsretval
\pgfplotstablegetelem{0}{[index]1}\of{\datatableM}  
\let\yfirstM\pgfplotsretval
\node[red, anchor=west, font=\scriptsize] at (axis cs:\xfirstM-10, \yfirstM) {mean};

\end{axis}
\end{tikzpicture}
    \caption{
    Final INL\textsubscript{max} estimation error as a function of the termination threshold $\varepsilon$, using the criterion described in \autoref{secsub:TerminationStrategy} with $N_\mathrm{term} = 12$, for a 16-bit ADC with $M=128$ samples per sweep and $1.0\,\mathrm{LSB}$ RMS noise.
    Each point corresponds to the result of $100$ independent runs, where a typical run (mean or 10th/90th percentile) terminates after a number of iterations determined by $\varepsilon$, yielding the corresponding INL error.
    For selected $\varepsilon$ values, bands indicate the full spread of errors across all 100 runs.
    }
    \label{fig:Epsilon}
\end{figure}

\subsection{Epsilon threshold tuning for termination}
\label{secsub:Results_Epsilon}
Using the previously selected inflation parameters, the termination behavior was evaluated for varying thresholds $\varepsilon$.
Based on a separate study, $N_\mathrm{term} = 12$ was chosen as it reliably avoided premature termination without extending test time unnecessarily.

To illustrate the behavior, the band for $\varepsilon = 10^{-4}$ is exemplary: the best $10\%$ of runs terminate in under $550$ iterations with an INL deviation lower than $0.026\,\mathrm{LSB}$, while the average result lies at $560$ iterations with $0.05\,\mathrm{LSB}$ error.
However, the slowest $10\%$ of runs require up to $720$ iterations, while the corresponding estimation error still remains below $0.1\,\mathrm{LSB}$.

In contrast, $\varepsilon = 10^{-3}$ (see \autoref{fig:Epsilon} orange band) reduces the required iteration count significantly to a range of $200-250$ iterations, while still keeping the estimation error below $0.07\,\mathrm{LSB}$ in $90\%$ of cases.
It therefore provides a favorable balance between runtime and accuracy.

A cross-comparison of \autoref{fig:Epsilon} and \autoref{fig:VarIter_Innovation} shows that the 90th percentile of the new inflation strategy reaches the same accuracy at iteration 220 as the 10th percentile of the original strategy at iteration 870. This implies a $4\times$ reduction in iteration count.
When combined with the $2\times$ speedup from the rank-1 update (for 16-bit ADCs), the overall test time to reach comparable accuracy is reduced by a factor of 8.

\subsection{Recommended parameter configuration}
\label{secsub:Results_Recommendation}
Based on the extensive simulations supporting the experimental results, the following parameters were found to be effective for practical use.
Each was chosen to balance convergence speed, estimation accuracy, and robustness.

\begin{itemize}
    \item Inflation factor: $\alpha = 9.1$
    \item NIS threshold: $\tau = 0.04$
    \item Termination threshold: $\varepsilon = 10^{-3}$
    \item Consecutive trace condition: $N_\mathrm{term} = 12$
\end{itemize}

In a practical test setup using an ADC operating at $1\,\mathrm{MS/s}$, the sample acquisition phase for each local sweep takes either $64\,\mu \mathrm{s}$ or $128\,\mu \mathrm{s}$ for $M=64$ or $M=128$ samples, respectively.
This window can be utilized to compute the next sample location in the asynchronous update loop.
The choice between $M = 64$ and $M = 128$ depends on the expected computation time per iteration, with $M = 128$ offering more time for processing.
Choosing $M$ as a power of two simplifies high-resolution DAC code spacing.

\section{Conclusion}
\label{sec:Conclusion}
This paper presented three significant enhancements to the Uncertainty-Guided Live Measurement Sequencing (UGLMS) method for real-time SAR ADC linearity testing.
(i) a rank-1 EKF update that cuts per-iteration complexity,
(ii) a low-order carrier polynomial extension to capture systematic nonlinearities, and
(iii) a measurement-aligned covariance-inflation strategy combined with a trace-based termination criterion for convergence-adaptive runtimes of individual test runs.
These refinements achieve equal-accuracy INL/DNL estimation $8\times$ faster for 16-bit ADCs and extend modeling and autonomy without offline post-processing.

Extensive simulations confirmed sub-$0.2\,\mathrm{LSB}$ accuracy even at 18 bit, completing full INL/DNL reconstruction in under $70\,\mathrm{ms}$ ($120\,\mathrm{ms}$ with the polynomial extension).
However, for lower resolutions (10–12 bit), the fixed overhead of the rank-1 update can outweigh its computational savings, making the original UGLMS more efficient in those cases, whereas the covariance-inflation strategy improves convergence at all resolutions.
Based on these simulations, a systematic parameter study identified recommended settings, chosen with typical real-time test constraints in mind.

Future work will focus on validating the proposed method through hardware measurements.

\section*{Acknowledgements}
This research was supported by Advantest as part of the Graduate School “Intelligent Methods for Test and Reliability” (GS-IMTR) at University of Stuttgart.

\section*{Declaration}
During the preparation of this work, the authors used ChatGPT 4o and Grammarly in order to polish parts of text for better readability. After using these tools/services, the authors reviewed and edited the content as needed and take full responsibility for the content of the publication.

\bibliographystyle{IEEEtran}
\bibliography{literature}

\end{document}